# Resource reduction for simultaneous generation of two types of continuous variable nonclassical states

Long Tian,[1, 2] Shaoping Shi,[1] Yuhang Tian,[1] Yajun Wang,[1, 2] Yaohui Zheng,[1, 2] and Kunchi Peng[1, 2]

[1]*State Key Laboratory of Quantum Optics and Quantum Optics Devices, Institute of Opto-Electronics, Shanxi University, Taiyuan 030006, China*
[2]*Collaborative Innovation Center of Extreme Optics, Shanxi University, Taiyuan, Shanxi 030006, China*

We demonstrate experimentally the simultaneous generation and detection of two types of continuous variable nonclassical states from one type-0 phase-matching optical parametric amplification (OPA) and subsequent two ring filter cavities (RFCs). The output field of the OPA includes the baseband $\omega_0$ and sideband modes $\omega_0 \pm n\omega_f$ subjects to the cavity resonance condition, which are separated by two cascaded RFCs. The first RFC resonates with half the pump wavelength $\omega_0$ and the transmitted baseband component is a squeezed state. The reflected fields of the first RFC, including the sideband modes $\omega_0 \pm \omega_f$, are separated by the second RFC, construct Einstein-Podolsky-Rosen entangled state. All freedoms, including the filter cavities for sideband separation and relative phases for the measurements of these sidebands, are actively stabilized. The noise variance of squeezed states is 10.2 dB below the shot noise limit (SNL), the correlation variances of both quadrature amplitude-sum and quadrature phase-difference for the entanglement state are 10.0 dB below the corresponding SNL.

## 1. INTRODUCTION

Continuous variable (CV) nonclassical states, including squeezed and Einstein-Podolsky-Rosen (EPR) entangled states, have been proven to be valuable resources in quantum information technology and have spread a variety of applications including quantum communication [1-7], quantum metrology [8-12], ultrahigh precision sensing [13-16] and quantum computer [17-19] etc. As central to the quantum system, nonclassical states generation is crucial to improve the quantum system performance. Intensive researches have focused on the improvement of squeezed or entanglement degree [20-27]. On the basis, it is fascinating to generate two types of nonclassical states simultaneously from one optical parametric amplification (OPA) without degrading squeezed or entanglement degree.

The optical parametric process has been proven as one of the most successful systems for nonclassical states generation, especially for the generation of high-level squeezed states [28-31], which has continually held the record for the largest amount of quantum noise reduction [32]. In such process, a high energy pump photon is converted into a pair of lower energy photons subject to energy conservation and the cavity resonance condition, via a second order nonlinear interaction [33-40]. The generation of a photon pair initiates a nonclassical correlation between the symmetric cavity modes along half the pump frequency. For a type-0 phase-matching parametric process, the downconversion modes with near-degenerate frequency and degenerate polarization are inseparable in space (squeezed state) [41]. The nonclassical noise correlation makes the noise variance of the photon pair beyond the shot noise limit (SNL). The downconversion modes from a type-II phase-matching parametric process with mutually orthogonal polarization can be conveniently separated by using a polarization beam splitter (PBS) [3], which presents nonclassical noise correlation, named as EPR entangled state. The different phase-matching requirement is a main obstacle of simultaneous generation of two types of nonclassical state from one OPA. Encouragingly, several types of phase-matching have been simultaneously realized in a single periodically poled KTiOPO$_4$ (PPKTP) crystal [42], and then the frequency up- and down-conversion processes were experimentally demonstrated at different operating temperature of the PPKTP [43, 44]. The switching operation was performed by manipulating the crystal temperature and the polarization direction of the pump and seed beams [44]. By time sharing multitask idea, two types of nonclassical states were successively obtained from one OPA.

Here, we present an experimental demonstration of generating simultaneously squeezed state (quantum fluctuation of 10.2 dB below the SNL) and EPR entangled state (quadrature amplitude-sum and quadrature phase-difference of 10.0 dB and 10.0 dB below the SNL) from one OPA. The output field of the OPA is spatially separated by employing two ring filter cavities (RFCs) as frequency-dependent beam splitter, which effectively saves quantum resource. The baseband around half the pump wavelength within the linewidth of the OPA is squeezed states. The sideband modes $\omega_0 \pm \omega_f$ construct EPR entangled states. The robust extension of the generation of two types of high performance CV nonclassical states from time sharing operation to simultaneous one will certainly promote their versatile applications in quantum information technology.

## 2. EXPERIMENTAL SETUP

A schematic of our experimental setup is illustrated in Fig. 1. The laser source of our experiment is a home-made single-frequency Nd:YVO$_4$ laser at 1064 nm with the output power of 2.5 W. The laser preparations, including spatial-mode improvement, intensity noise suppression, polarization purification and second harmonic generation, are similar to our earlier experiments presented in Refs [29, 45, 46]. Our OPA is a semi-monolithic cavity consisting of a piezoactuated concave mirror and a PPKTP crystal with the dimensions of $1\text{mm} \times 2\text{mm} \times 10\text{mm}$. The crystal end face with a radius of curvature of 12 mm is coated as high reflectivity (HR) for the fundamental field, thus serving as the cavity end mirror. The concave mirror with a radius of curvature of 30 mm has a transmissivity of 12% ±1.5% for 1064 nm. The air gap between the crystal and the coupling mirror is 27 mm, corresponding to a free spectral range (FSR) $\omega_f$ of 3.328 GHz. The phase-matching bandwidth of the PPKTP crystal is approximately 2 THz, corresponding to more than 300 pairs of downconversion modes subject to cavity resonance condition.

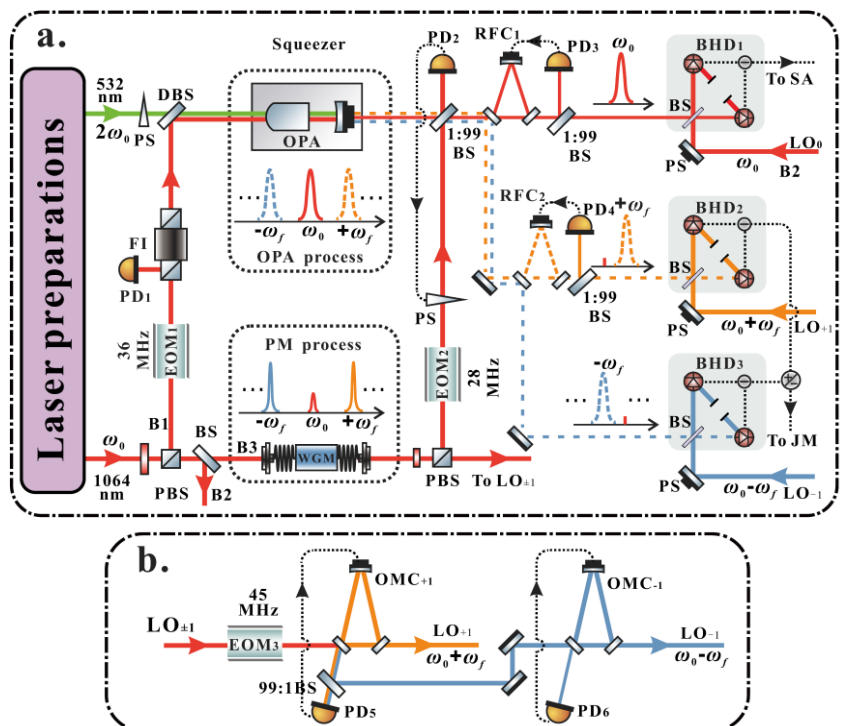

Fig. 1 Schematic of the experimental setup. OPA, optical parametric amplifier; PM, phase modulation; WGM, waveguide electro-optic modulator; EOM$_{1-3}$, electro-optic phase modulator; OMC$_{\pm 1}$, optical mode cleaner; RFC$_{1-2}$, ring filter cavity; FI, faraday isolator; PS, phase shifter; DBS, dichroic beam splitter; PBS, polarized beam splitter; BS, beam splitter; BHD$_{1-3}$, balanced homodyne detector; PD$_{1-6}$, photodetector; $\omega_f$, free spectral range of the OPA; SA, spectrum analyzer; JM, joint measurement.

The laser beam at 532 nm serves as the pump field of the OPA and does not resonant with the OPA, which has good immunity to temperature variation. The laser at 1064 nm is divided into three beams B1, B2 and B3. The B1, which is modulated by $EOM_1$ with frequency of 36 MHz, serves as the initial alignment beam of OPA and the seed beam for stability of the OPA cavity length and relative phase between pump and seed beams. We use $PD_1$ to collect the reflected beam of OPA for extracting these error signals. The beam B2 is employed as the local oscillator (LO) for squeezed noise detection at baseband $\omega_0$.

The beam B3 passes through the WGM with a modulated frequency of $\omega_f$, and its output includes the carrier frequency $\omega_0$ and higher harmonics $\omega_0 \pm \omega_f$. The power of each sideband mode produced by the WGM is large enough to generate error signals for downstream experiment. Small part of the WGM output is split by a PBS and passes through an optical mode cleaner ($OMC_{+1}$) immediately to get rid of the unwanted carrier and transmitted corresponding upper sideband components $LO_{+1}$. The linewidth (FSR) of the OMC is 2 MHz (1.29 GHz), which ensures that about $10^{-5}$ of the other sideband modes transmits through the OMC after the cavity resonating with one of the sidebands. The reflection light of $OMC_{+1}$ passes through another $OMC_{-1}$ to extract the corresponding low sideband components $LO_{-1}$. Then, the frequency-shifted $LO_{\pm 1}$ with frequency of $\omega_0 \pm \omega_f$ are generated for further quadrature measurements.

The beam, emitting out from the OPA, includes many pairs of downconversion modes subject to the cavity resonance condition within the spectral acceptance bandwidth of the PPKTP. Besides that, the output of the OPA is bright squeezed light with coherent amplitude and can be used as a sensor to lock the $RFC_1$ to resonate with the baseband mode of squeezed light. However, we cannot extract any signals needed for actively stabilizing $RFC_2$ resonance at a required sideband mode owing to these sideband modes of squeezed light are vacuum fields without coherent amplitude. The relative phase between a given sideband and the corresponding $LO_0$ in the downstream experiment is also out of control due to the same dilemma. For resolving the problem, we use the rest component of B3 as the auxiliary beam for stabilizing the $RFC_2$ length and the relative phase between the sideband modes and frequency-shifted $LO_{\pm 1}$.

The auxiliary beam is modulated by $EOM_2$ with frequency of 28 MHz and coupled with the bright squeezed beam on a 1:99 beam-splitter (BS). Then, the $PD_2$ collects the one port of the coupled light to extract the error signal for stably locking the carrier of the auxiliary beam to that of the squeezed beam to zero. Finally, the auxiliary beam, including its sideband modes $\omega_0 \pm \omega_f$, can be adopted for controlling the length of downstream RFCs and relative phase stability of balanced homodyne detectors (BHD) instead of vacuum sideband modes. Then, we can use the auxiliary beam to lock the two RFCs to resonate respectively with the sideband modes $\omega_0 \pm \omega_f$ of the bright squeezed light.

A cascade of two RFCs is employed as the band filters to carry out the sideband modes separation. Two RFCs resonate respectively with the baseband mode $\omega_0$ and sideband modes $\omega_0 + \omega_f$, and thus the different frequency components are transmitted from each RFC. The RFC consists of two plane mirrors and one curved mirror with the round-trip length of 210 mm, and the curved cavity mirror shows a transmissivity of 0.008% with the curvature radius of 1 m. The two plane cavity mirrors have a power transmissivity of 13.5% for s-polarized beam. These provide finesse value of 21.6 and linewidth of 66 MHz for s-polarized beam, in accordance with the round-trip length. These parameters can ensure that the resonance mode has a high transmission and the neighboring sideband modes are effectively reflected.

The transmitted field $\omega_0$ (squeezed state) from the $RFC_1$ is interfered with the $LO_0$ at the frequency $\omega_0$ to detect the quadrature noise variances at baseband. The transmitted and reflected fields from the $RFC_2$ (entanglement state) are interfered with the $LO_{+1}$ and $LO_{-1}$, respectively, to quantify the quantum correlations between the sideband modes $\omega_0 \pm \omega_f$. We split 1% of the coherent sideband modes at the outputs of these RFCs and the reflected fields of OMCs to stabilize their cavity lengths, respectively. The error signals are extracted by resonant photodetectors with high Q value. After corresponding beam splitter (BS), the output beams are coupled toward BHDs with a particular phase control branch to detect the noise variance of the quadrature components. The detector is built from a pair of p-i-n photodiode with the photo surface of 100 μm. To recycle the residual reflection from photodiode surfaces, three pairs of concave mirrors with the curvature radius of 50 mm are used as retroreflectors.

## 3. EXPERIMENTAL RESULTS

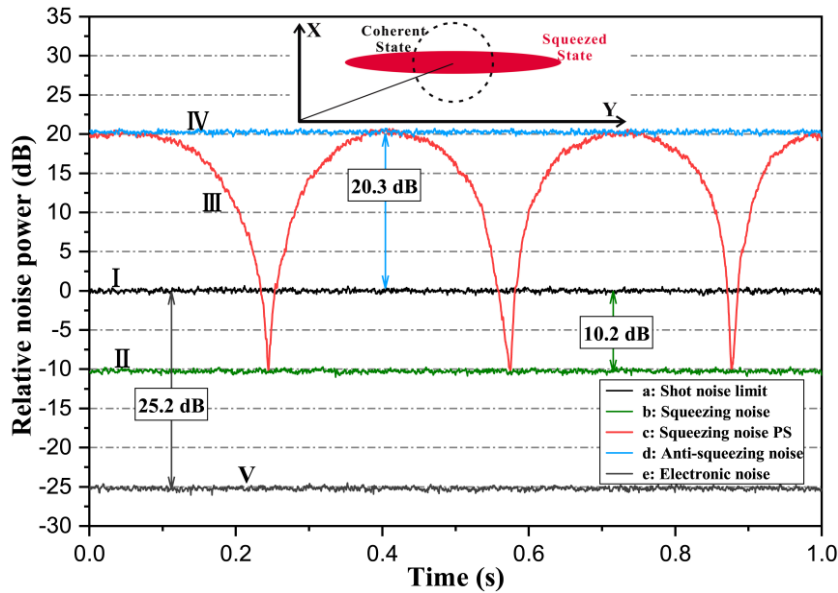

Fig. 2 Quantum noise levels of the squeezed and anti-squeezed state are recorded at the analysis frequency of 2 MHz with resolution bandwidth (RBW) 300 kHz and video bandwidth (VBW) 200 Hz. These data still include electronic noise, and represent direct observations.

Locking the relative phase between the pump and seed beam to π, the OPA operates at deamplification. The RFC$_1$ resonating with half the pump wavelength $\omega_0$ transmits the baseband mode of the bright squeezed light. Figure 2 presents the measured results of the transmitted baseband mode (squeezed state) at the pump power of 150 mW. All traces are measured by a spectrum analyzer (Agilent N9020A with the uncertainty of 0.2 dB). Trace I corresponds to the shot noise limit of 5.5 mW LO$_0$ power and is measured with the squeezed light blocked. Trace II shows the quantum noise reduction when squeezed states are injected with the relative phase 0 between the signal beam and LO$_0$. The directly observed squeezing level is 10.2 dB at the analysis frequency of 2 MHz without subtracting electronic noise. Trace III is a variational noise level with the local oscillator phase scanned. Trace IV is a noise level with the relative phase π/2, corresponding to the anti-squeezing of 20.3 dB above the SNL. Trace V is electronic noise, which is 25.2 dB below the SNL at the LO$_0$ power of 5.5 mW.

The RFC$_2$, resonating with the sideband mode $\omega_0 + \omega_f$, is located at the reflection end of the RFC$_1$. The transmitted and reflected sideband modes $\omega_0 \pm \omega_f$ are separated by the RFC$_2$, resonating with the sideband mode $\omega_0 + \omega_f$, constructing EPR entangled state. We perform the detection of quantum correlation between the two sideband modes $\omega_0 \pm \omega_f$. The correlation measurements of amplitude-sum and phase-difference between the upper and lower sidebands measured by BHDs are shown in Fig. 3. With the signal ports of the two BHDs blocked, the output of the power combiners corresponds to the SNL (trace I in Fig. 3). When the relative phases between the LO$_{\pm 1}$ and two sidebands of squeezes states are 0 (π/2), the outputs of two BHDs are combined with the upper (low) power combiners to obtain the noise powers of amplitude sum (phase-difference) shown in trace II in Fig. 3(a) (Fig. 3(b)). Either of the BHD's phase keeps constant while another BHD's phase is scanned linearly in time and the measured noise variances are shown in trace III in Fig. 3. When one of the BHDs signal port is blocked, the noise spectrum is higher than the SNL, which is insensitive to the relative phase (trace IV in Fig. 3(a) and Fig. 3(b)). The variances of amplitude sum and phase difference of the EPR sideband entanglement are both 10.0 dB, without the electronic noise subtracted.

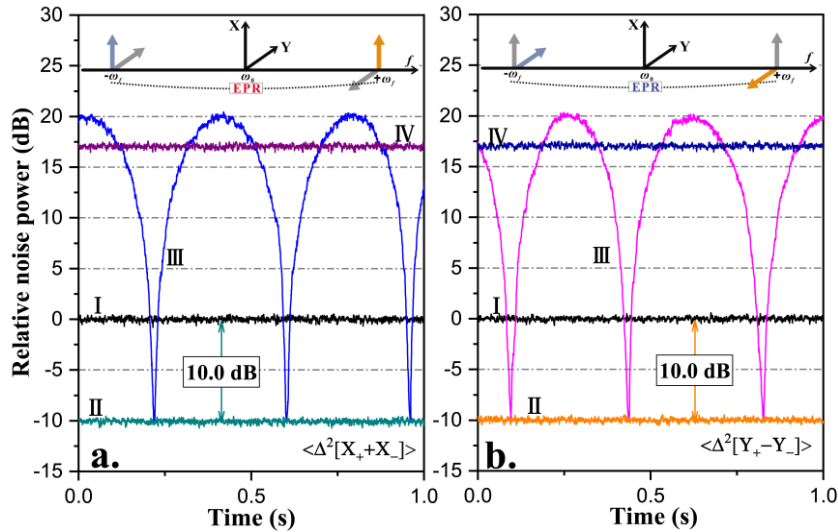

Fig. 3 Normalized correlation noise of amplitude-sum (a) and phase-difference (b) between the upper and lower sidebands $\omega_0 \pm \omega_f$. Trace I, shot noise limit; Trace II, correlation noise of $<\Delta^2[X_+ + X_-]>$ (a) and $<\Delta^2[Y_+ - Y_-]>$ (b); Trace III, correlation noise variance with one of the BHD's phases of 0 (a) and $\pi/2$ (b) while another BHD's phase is scanned linearly in time; Trace IV, noise spectrum with one of the BHD signal port is blocked.

For our generation scheme of entanglement state, the two sidebands of squeezed states are Gaussian state and strictly symmetric, which can be verified by the Duan inseparability criterion in the following form [47]:

$$Var(\hat{X}_A + \hat{X}_B) + Var(\hat{Y}_A - \hat{Y}_B) < 2 \qquad (1)$$

Where *Var* denotes the variance, with the variance of a vacuum field normalized to unity. $\hat{X}_A$ and $\hat{X}_B$ are the quadrature amplitude operators. $\hat{Y}_A$ and $\hat{Y}_B$ are the quadrature phase operators orthogonal to $\hat{X}_A$ and $\hat{X}_B$, respectively. All the generated EPR entangled beams above satisfy the inseparable criteria $Var(\hat{X}_A + \hat{X}_B) + Var(\hat{Y}_A - \hat{Y}_B) = 0.20 < 2$.

## 4. CONCLUSION

In conclusion, we demonstrate experimentally the simultaneous generation and detection of two types of nonclassical states by a type-0 phase-matching OPA and subsequent two RFCs. The scheme effectively saves quantum resource. The output field of the OPA includes the baseband $\omega_0$ and sideband modes $\omega_0 \pm \omega_f$ subject to the cavity resonance condition. The RFC$_1$ resonates with half the pump wavelength $\omega_0$ and transmits the baseband mode of the OPA. The baseband includes these symmetric downconversion modes along half the pump wavelength within the linewidth of the OPA, which presents noise variance below the SNL (squeezed state). The sideband modes $\omega_0 \pm \omega_f$ reflected from the RFC$_1$ are spatially separated by the RFC$_2$, constructing EPR entangled state. Both of the correlation noises of squeezed and entanglement states are superior to 10 dB below the SNL. In virtue of the auxiliary beam generated by a WGM, all freedoms in downstream experiment are actively stabilized. As central to the CV quantum system, the simultaneous generation of squeezed and entanglement states will promote a practical and versatile requirement of quantum information technology.

**Acknowledgements** This work was supported by the National Natural Science Foundation of China (11654002, 11804207, 11874250, 11804206); National Key Research and Development Program of China (2016YFA0301401); Program for Sanjin Scholar of Shanxi Province; Key Research and Development Program of Shanxi (201903D111001); Fund for Shanxi 1331 Project Key Subjects Construction; Program for Outstanding Innovative Teams of Higher Learning Institutions of Shanxi; Natural Science Foundation of Shanxi Province (201801D221006).